\documentclass[twocolumn,showpacs,amsmath,amssymb,prb]{revtex4-1}
\usepackage{graphicx}
\usepackage{dcolumn}
\usepackage{bm}
\begin{document}

\title{Orbital order and fluctuations in the two-leg ladder materials BaFe$_2X_3$ ($X$ = S and Se) and CsFe$_2$Se$_3$}

\author{Kou Takubo}
\email{ktakubo@issp.u-tokyo.ac.jp}
\affiliation{Institute for Solid State Physics, University of Tokyo, Kashiwa, Chiba 277-8581, Japan}
\author{Yuichi Yokoyama}
\affiliation{Institute for Solid State Physics, University of Tokyo, Kashiwa, Chiba 277-8581, Japan}
\author{Hiroki Wadati}
\affiliation{Institute for Solid State Physics, University of Tokyo, Kashiwa, Chiba 277-8581, Japan}
\author{Shun Iwasaki}
\affiliation{Department of Applied Physics, Waseda University, Okubo 169-8555, Japan}
\author{Takashi Mizokawa}
\affiliation{Department of Applied Physics, Waseda University, Okubo 169-8555 Japan}
\author{Teak Boyko}
\affiliation{Canadian Light Source, Saskatoon, Saskatchewan S7N 2V3, Canada}
\author{Ronny Sutarto}
\affiliation{Canadian Light Source, Saskatoon, Saskatchewan S7N 2V3, Canada} 
\author{Feizhou He}
\affiliation{Canadian Light Source, Saskatoon, Saskatchewan S7N 2V3, Canada}
\author{Kazuki Hashizume}
\affiliation{Department of Physics, Tohoku University, Sendai 980-8578, Japan}
\author{Satoshi Imaizumi}
\affiliation{Department of Physics, Tohoku University, Sendai 980-8578, Japan}
\author{Takuya Aoyama}
\affiliation{Department of Physics, Tohoku University, Sendai 980-8578, Japan}
\author{Yoshinori Imai}
\affiliation{Department of Physics, Tohoku University, Sendai 980-8578, Japan}
\author{Kenya Ohgushi}
\affiliation{Department of Physics, Tohoku University, Sendai 980-8578, Japan}

\date{\today}

\begin{abstract}

The electronic structure of BaFe$_2X_3$ ($X$ = S and Se) and CsFe$_2$Se$_3$ in which two-leg ladders are formed by the Fe sites are studied by means of x-ray absorption and resonant inelastic x-ray scattering spectroscopy.
The x-ray absorption spectra at the Fe $L$ edges for BaFe$_2X_3$ exhibit two components, indicating that itinerant and localized Fe 3$d$ sites coexist.
Substantial x-ray linear dichroism is observed in polarization dependent spectra, indicating the existence of orbital order or fluctuation in the Fe-ladder even above the N\'{e}el temperature $T_{\rm N}$.
Direct exchange interaction along the legs of the Fe-ladder stabilizes the orbital and antiferromagnetic orders in BaFe$_2$S$_3$, while the ferromagnetic molecular orbitals are realized between the rungs in CsFe$_2$Se$_3$.

\end{abstract}

\pacs{78.70.Dm, 78.70.En, 74.70.Xa, 75.25.Dk }
\maketitle

\section{Introduction}
The magnetic-orbital fluctuations and their anisotropies in iron-based superconductors have been attracting much attention.
The parent compounds of the iron-based superconductors show antiferromagnetic (AF) transitions at low temperatures, typically exhibiting striped-type magnetic ordering.\cite{Cruz08,Huang08,Dai15}
On the basis of theoretical analyses on multiband models with hole and electron Fermi pockets, the striped-type magnetic ordering is stabilized by Fermi-surface nesting, and the associated AF and orbital fluctuations are proposed to induce the superconductivity.\cite{Mazin,Kuroki,Lee09,Dai15,Yamakawa,Chubukov}
However, there are some iron-based superconductors showing significant disagreement with the Fermi-surface nesting scenario.
For example, superconductivity with $T_{\rm c}\sim$30 K in 245 system ($A_2$Fe$_4$Se$_5$, $A$ = K, Rb, and Cs)\cite{Guo,Ye,Shermadini} appears in the vicinity of the Mott insulating state with block-type AF ordering.
In this context, it is very important to study the nature of the Mott insulating state in the parent compounds of iron-based superconductors.

Recently, another insulating Fe chalcogenide $A$Fe$_2X_3$ ($A$ = Cs and Ba, $X$ = S and Se) has been attracting attention due to the specific quasi-one-dimensional crystal structure and magnetism.\cite{Hong,Krzton,Caron11,Caron12,Nambu,Du,Chi,Takahashi,Yamauchi,Lei,Hirata,Monney,Ootsuki,Popovic,Luo,Dong}
In this family of compounds, Fe(S,Se)$_4$ tetrahedra share their edges and form a two-leg ladder of Fe sites as shown in Fig. 1 (a).
These compounds all exhibit unique magnetic ordering.
The magnetic structure of BaFe$_2$Se$_3$ ($Pnma$ space group) is a one-dimensional analog of the block magnetism observed in $A_2$Fe$_4$Se$_5$, in that four Fe spins in the two-leg ladder form a ferromagnetic block and the neighboring blocks are antiferromagnetically coupled as illustrated in Fig. 1 (f).\cite{Krzton,Caron11,Caron12,Nambu}
In contrast, the magnetic structures of BaFe$_2$S$_3$ and CsFe$_2$Se$_3$ ($Cmcm$ space group), are of the stripe-type, in which the magnetic moments couple ferromagnetically along the rung, and antiferromagnetically along the leg direction.\cite{Du,Chi}
However, the magnetic moments in CsFe$_2$Se$_3$ point toward the layers, while those in BaFe$_2$S$_3$ point toward the rungs.

\begin{figure}[t!]
	\includegraphics[width=1\linewidth]{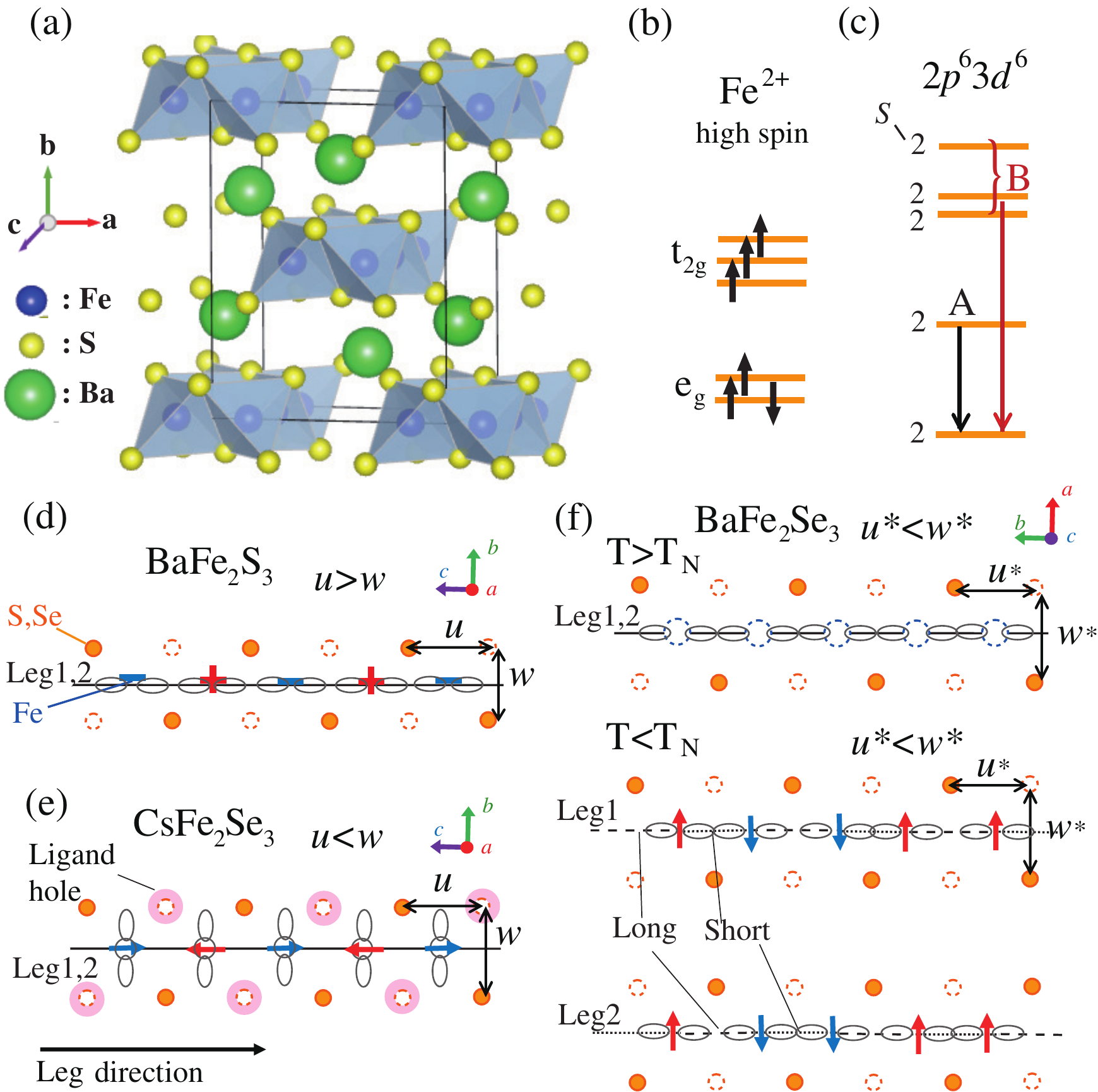}%
	\caption{(Color online)
		Crystal and electronic structure of BaFe$_2X_3$ and CsFe$_2$Se$_3$.
		(a) Crystal structure of BaFe$_2$S$_3$ visualized using the software package Vesta.\cite{Homma}
		(b) Electronic configuration of Fe$^{2+}$ high-spin state in a tetrahedral symmetry.
		(c) Multiplet levels of the	initial state for $2p^63d^6$ suggested in Ref. \onlinecite{Monney}.
		The Coulomb and crystal field interactions are considered with associated $S$ number.
		A and B mark the two dominant \mbox{$d$-$d$} excitation peaks seen in the RIXS experiment.
		(d),(e),(f) Schematic drawing of the magnetic structure and lattice distortion in a leg of Fe ladder for (d) BaFe$_2$S$_3$, (e) CsFe$_2$Se$_3$, and (f) BaFe$_2$Se$_3$, viewed from a side of the ladder.
		The definitions of crystal axes are different between BaFe$_2$S$_3$, CsFe$_2$Se$_3$, and BaFe$_2$Se$_3$.
		The S or Se layers locate up (solid circle) and down (dotted circle) the Fe-leg layer and consist of Fe(S,Se)$_4$ tetrahedra. 
		Gray lobes indicate examples of occupied orbitals.
		Magenta shades in (e) indicate the ligand holes suggested in Ref. \onlinecite{Ootsuki}. 
	}
\end{figure}

Moreover, recent works under high-pressures revealed that BaFe$_2$S$_3$, which is the most conductive compound among these compounds, shows an appearance of the superconducting phase at $T_{\rm c}$$\sim$14 K under 11 GPa without major crystal structure change.\cite{Takahashi}
Also it has been clarified that the superconducting phase is in the vicinity of bandwidth-control type Mott transition.\cite{Yamauchi}
In addition, the magnetic transition temperature and resistivity of BaFe$_2X_3$ (Fe$^{2+}$) depend on the sample stoichiometry.
Lei \textit{et al}. reported the activation-type temperature dependence for BaFe$_2$Se$_3$ with a band gap of 0.18 eV.\cite{Lei}
On the other hand, one-dimensional variable range hopping was reported, indicating that some carriers are localized due to strong scattering effects in the quasi-one-dimensional structure.\cite{Nambu,Yamauchi,Hirata}
Moreover, coexistence of the itinerant and localized electrons was indicated by the resonant inelastic x-ray scattering (RIXS),\cite{Monney} and x-ray photoemission spectroscopy (XPS).\cite{Ootsuki}
These observations suggest that the itinerant electrons introduced by small Fe vacancies or some other effects would be responsible for the variable range-hopping behavior of the resistivity.\cite{Nambu,Hirata,Yamauchi}
In contrast to BaFe$_2X_3$, CsFe$_2$Se$_3$ with formal Fe valence of +2.5 is much more insulating.
Usually, Mott insulators with integer number of valence are expected to be more insulating than the mixed valence systems.
Such puzzling mismatch between the formal valence and the transport behavior indicates unusual electronic states in the vicinity of the superconducting phase of $A$Fe$_2X_3$.
The x-ray linear dichroism (XLD) for x-ray absorption spectroscopy (XAS) and RIXS are ideal tools to detect electronic anisotropy in such systems with charge and orbital degrees of freedom of transition-metal 3$d$ electrons.

In other families of the iron-based superconductors, 122 ($A$Fe$_2$As$_2$, $A$ = alkaline-earth) and 1111 ($R$FeAsO, $R$ = rare earth) systems, 
the N\'{e}el temperature ($T_{\rm N}$) and structural transitions ($T_{\rm s}$) are split, which have recently been considered as a manifestation of electronic nematic order.\cite{Fernandes14}
These orders have been inferred from the unusual anisotropy in resistivity,\cite{Chu10,Ying11,Chu12} optical conductivity,\cite{Dusza11} and orbital occupancy\cite{Yi11,Kim} observed at the temperatures above $T_{\rm s}$ and $T_{\rm N}$. 
In the present paper, we investigate the electronic structure of
BaFe$_2$S$_3$, BaFe$_2$Se$_3$, and CsFe$_2$Se$_3$ in the Fe sites by means of XAS and RIXS at the Fe $L_{2,3}$ absorption edges.
An opposite XLD, namely the electronic anisotropy, is observed for BaFe$_2X_3$ and CsFe$_2$Se$_3$ at room temperature, indicating the existence of the orbital order or fluctuation above $T_{\rm N}$.
The orbital and AF order along the legs of Fe ladder is emerged via the direct exchange interaction between the Fe sites in BaFe$_2$S$_3$. 
On the other hand, the molecular orbital formation along the rung is associated in CsFe$_2$Se$_3$.

\section{Experiment}
Single crystals of BaFe$_2$S$_3$, BaFe$_2$Se$_3$, and CsFe$_2$Se$_3$ were grown by the melt-growth method.\cite{Hirata,Du}
XAS and RIXS measurements were performed at the REIXS beamline of the Canadian Light Source.\cite{Hawthorn}
The single crystals were cleaved at room temperature (300 K) under the base pressure of 5$\times$10$^{-6}$ Pa for the XAS and RIXS measurements.
The cleaved surfaces were oriented to the (110) planes for BaFe$_2$S$_3$ and CsFe$_2$Se$_3$, and (100) plane for BaFe$_2$Se$_3$, parallel to the legs of ladder.
Although the crystals of BaFe$_2$Se$_3$ consist of some blocks misaligned by a rotation along the ladder direction,\cite{Krzton,Monney} this fact does not seriously affect the main conclusion of XLD for $E//{\rm leg}$ or $E\perp{\rm leg}$ discussed later.
The XAS spectra were recorded both in the total-electron-yield (TEY) and total-fluorescence-yield (TFY) modes.
At the RIXS measurement, the samples were measured at the incident angle of 60$^\circ$ and the emissions were detected at $\theta$=90$^\circ$ for the x ray [see inset of Fig. 5 (d) about the experimental geometry].
The beamline slit was set to 25 ${\mu}$m, which resulted in an effective combined resolution of both the incoming beam and spectrometer of $\sim$0.8 eV for RIXS measurements at the Fe $L_3$ edge.
The energy of outgoing photons was calibrated by a reflection from a copper plate.

\section{Results and Discussion}
Figure 2 shows the XAS spectra at the Fe $L_{2,3}$ absorption edge of BaFe$_2$S$_3$, BaFe$_2$Se$_3$, and CsFe$_2$Se$_3$ taken with the (a) TEY and (b) TFY modes at room-temperature.
Spectral difference between the less-distorted TEY and bulk-sensitive TFY spectra is barely observed, indicating the clean ordered surface for these samples. 
The two white lines in the spectra result from 2$p$ to 3$d$ dipole transitions ($2p^63d^6\rightarrow 2p^53d^7$) with the well-separated spin-orbit-split 2$p$ states $2p_{3/2}$ ($L_3$) and $2p_{1/2}$ ($L_2$), appearing respectively at around $\sim$708 and $\sim$721 eV.
No sharp multiplet is observed in the spectra, that exhibits a similar spectral shape as 
chalcogenides of Fe(Se,Te)\cite{Saini} and Fe-pnictide materials.\cite{Yang,ParksCheney,Bondino08,Bondino10,Kim}
Additionally, some spectral weights can be seen in the energy range of 3-5 eV above the white lines.
These features are an indication of interaction of the chalcogen sites with the Fe 3$d$ states\cite{Saini} and is notably well-separated in the sharp spectrum for CsFe$_2$Se$_3$, which are indicated in the red arrows in Fig. 2(a).
Despite the formal Fe valence of +2.5 for CsFe$_2$Se$_3$, the spectrum of CsFe$_2$Se$_3$ is very sharp but consistent with the observation of a previous Fe 2$p$ XPS study,\cite{Ootsuki} indicating a localized Mott insulating nature with Fe$^{2+}$.
If all the Fe sites in CsFe$_2$Se$_3$ take the high-spin Fe$^{2+}$ configuration, the extra positive charge (+0.5 per Fe) should be located at the Se sites.
Since the charge-transfer energy from Se 4$p$ to Fe 3$d$ is small, 
if Fe$^{3+}$ exists in CsFe$_2$Se$_3$, it should take the $d^6\underline{L}$ configuration instead of $d^5$.\cite{Ootsuki}
In this ligand hole picture, the two-leg ladder in CsFe$_2$Se$_3$ accommodates the $d^6$-like and $d^6\underline{L}$-like sites.
Assuming that the $d^6$-like and $d^6\underline{L}$-like sites are aligned along the rung, the Se 4$p$ hole should be located at the Se sites sandwiched by the two legs as schematically shown in Fig. 1(e).

\begin{figure}[t!]
	\includegraphics[width=1\linewidth]{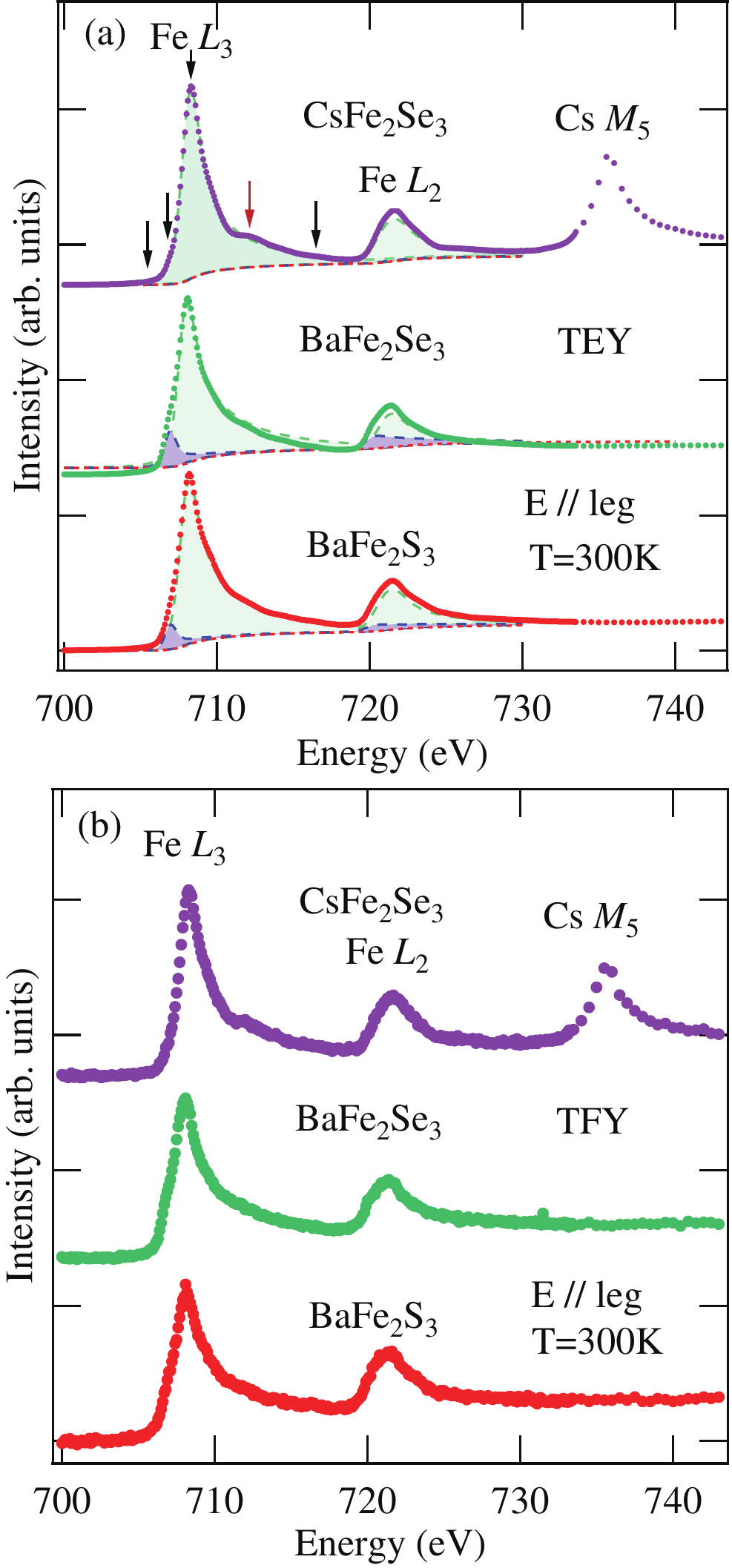}%
	\caption{(Color online) 
		XAS spectra at the Fe $L_{2,3}$ edges for BaFe$_2X_3$ ($X$ = S and Se) and CsFe$_2$Se$_3$ taken in the (a) TEY and (b) TFY modes.
		The dashed lines indicate the results of curve fitting.
		The arrows indicate the incident energies used in the RIXS measurements shown in Fig. 5.
	}
\end{figure}

On the other hand, some shoulder structures below the white lines are observed in the spectra of BaFe$_2$S$_3$ and BaFe$_2$Se$_3$, that are also similar to the Fe 2$p$ XPS spectra and corresponding to rather electron doping compared to the case of CsFe$_2$Se$_3$.
In Fig. 2(a), the results of Mahan's-line shape fitting are indicated by the dashed curves. 
The weak but significant components are observed at the pre-edge region $\sim$1.0 eV below the white lines for BaFe$_2$Se$_3$ and BaFe$_2$S$_3$.
On the XPS study, these two components have been ascribed as the contribution from itinerant and localized electrons.
Since the pre-edge region of XAS for transition-metal $L_{2,3}$ edges corresponds to the transition to unoccupied $d$ state near the Fermi-level,
these low energy structures originate from the itinerant $e_g$ empty state of the Fe$^{2+}$ high-spin state [see Fig.1 (b)].
The itinerant and localized electrons will coexist in BaFe$_2$Se$_3$ and BaFe$_2$S$_3$.
Namely, self-dopings from the Se 4$p$ and S 3$p$ to Fe 3$d$ states arise from smallness of the charge transfer and cause partial delocalization of the electrons in these systems. 
On the other hand, the Fe 3$d$ electrons with the Fe$^{2+}$ high-spin configuration and the Se 4$p$ holes are localized in CsFe$_2$Se$_3$.

In addition to these features, the XAS spectra at the Fe edges exhibit unique anisotropies.
Figure 3 gives the XLD spectra of BaFe$_2X_3$ ($X$ = S and Se) and CsFe$_2$Se$_3$ at room temperature.
The spectra are normalized by the area between 700 eV and 718 eV.
The substantial XLD ($I_{E//{\rm leg}}-I_{E\perp{\rm leg}}$) are observed for all samples and exhibit an opposite behavior for BaFe$_2X_3$ and CsFe$_2$Se$_3$.
The sign of XLD is minus for BaFe$_2$S$_3$ below the $L_3$ main peak of 708.1 eV in the overall pre-edge region and plus in the higher energy region.
XLD observed in the spectra for BaFe$_2$S$_3$ is fairly similar to that obtained for BaFe$_2$As$_2$ ($I_{AF}-I_{ferromagnetic}$) below $T_{\rm s}$.\cite{Kim}
XLD for BaFe$_2$As$_2$ was consistent with the existence of an orbital order along their AF direction and an opposite to tendency of their local structural distortion of $a_{AF}>b_{ferromagnetic}$.\cite{Kim,Chen}
The spectrum for BaFe$_2$S$_3$ in the pre-edge region taken with the $E//{\rm leg}$ ($E\perp{\rm leg}$) polarization detects the empty $e_g$ state of the $d_{3z^2-r^2}$ ($d_{x^2-y^2}$) orbital as discussed earlier.
Here, $z$ is defined to be parallel to the leg direction, or $c$ for BaFe$_2$S$_3$ and CsFe$_2$Se$_3$, and $b$ for BaFe$_2$Se$_3$, respectively.   
Therefore, $d_{3z^2-r^2}$ will be occupied and the orbital order along the leg direction  is indicated in BaFe$_2$S$_3$.

\begin{figure}[t!]
	\includegraphics[width=1\linewidth]{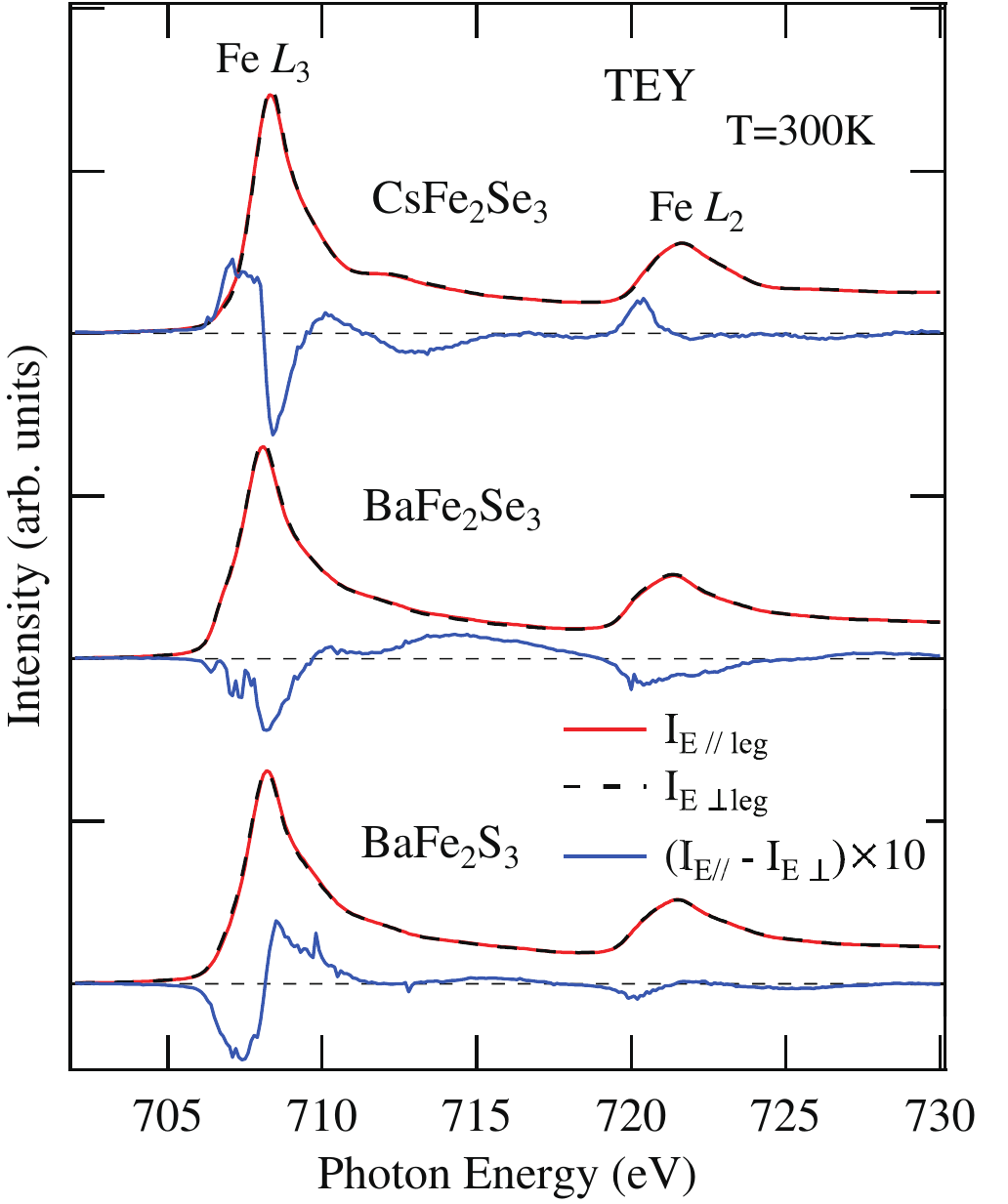}%
	\caption{(Color online)
		XAS spectra at the Fe $L_{2,3}$ edges with two different polarizations taken in the TEY modes at $T$ = 300 K for BaFe$_2X_3$ ($X$ = S and Se) and CsFe$_2$Se$_3$.
		The blue curves are the XLD spectra.		
	}
\end{figure}

In the case of BaFe$_2$Se$_3$, most of the $d_{3z^2-r^2}$ orbitals should be occupied,
since XLD in the pre-edge region of $\sim$707 eV is essentially similar to that of BaFe$_2$S$_3$.
However, a dip-hump like shape is observed at the pre-edge region and 
the sign of XLD is still minus at the main-peak region of $\sim$708 eV.  
Since the crystals of BaFe$_2$Se$_3$ has many twins at room temperature compared to BaFe$_2$S$_3$ and CsFe$_2$Se$_3$,\cite{Hirata}
it is possible that the crystal of BaFe$_2$Se$_3$ contains many defects.
The small Fe vacancy will affects the electronic configuration for BaFe$_2$Se$_3$ and the orbital order may become more complicated, corresponding to their block-type magnetism and lattice distortion below $T_{\rm N}$.

On the other hand, there is a peak in the pre-edge region of XLD for CsFe$_2$Se$_3$ and it exhibits a dip at the main-peak structure around 708.4 eV.
Therefore, more holes are suggested in the $d_{3z^2-r^2}$ orbital rather than in the $d_{x^2-y^2}$.
Namely, the orbital order perpendicular to the leg direction is indicated in CsFe$_2$Se$_3$, in contrast to BaFe$_2$S$_3$ [see also Fig. 1(e)].
However, the very sharp Fe peaks with the satellite-like structures of CsFe$_2$Se$_3$ cannot be simply understood for the formal valence of 2.5+ with any kind of the orbital order in the Fe sites, whereas this observation is still consistent with the insulating nature of this compound.

One may consider that the anisotropic spectra discussed above are naturally expected on the basis of their local structural distortions in the two-leg ladder of $A$Fe$_2$X$_3$, and do not link to the existence of orbital orders.
However, the present tendency of XLD for BaFe$_2$S$_3$ is opposite to its local distortion, which was theoretically clarified by Chen \textit{et al.} for BaFe$_2$As$_2$.\cite{Chen}
The local distortions around the Fe sites can be described by the elongation or compression for the FeS$_4$ tetrahedra.
The FeS$_4$ tetrahedra of BaFe$_2$S$_3$ are elongated along the leg direction ($u/w\sim$1.03)\cite{Hong} [see Fig. 1(d)], which corresponds to the AF direction for BaFe$_2$As$_2$.
In this case, the dip-like structure would be observed in XLD of the higher energy region and hump would be observed in the lower energy region of the Fe $L_3$ edge.\cite{Chen}
However, the observed XLD for BaFe$_2$S$_3$ is opposite to these tendencies.
On the other hand, the FeSe$_4$ tetrahedra of CsFe$_2$Se$_3$ are compressed ($u/w\sim$0.98) along the leg.\cite{Du}
Moreover, the orbital order in many iron-based superconductors including BaFe$_2$As$_2$ were unable to be reproduced by simple first-principle calculations based on the structural data and still has been a controversial problem.
Actually, a recent first-principle calculation for paramagnetic metallic state of BaFe$_2$S$_3$ indicates the different orbital filling in the Fe 3$d$ states ($n_{d_{x^2-y^2}}>n_{d_{3z^2-r^2}}$).\cite{Arita15,Suzuki15}
Therefore, the observed XLD for BaFe$_2$S$_3$ and CsFe$_2$Se$_3$ cannot be explained from their structural distortions, indicating the existence of orbital order or fluctuation at room temperature.

Furthermore, the tendency of XLD in the lower energy region of the Fe $L$ edges show scarcely any temperature dependence as shown in Fig. 4.
While the higher energy structures of $L_3$ around $\sim$709 eV show some cleavage dependences owing to the defects and/or contaminations, the tendency and magnitude of XLD are essentially similar to those obtained at Fig. 3.
As the temperature decreases, the magnitudes of XLD for all samples steadily increase but irrelevant to $T_{\rm N}$.
Those for BaFe$_2$S$_3$ and BaFe$_2$Se$_3$ at $T$=100 K (and $T$=200 K) are $\sim$ 30$\%$ (and $\sim$ 10$\%$) larger than those at $T$=300 K. 
On the other hand, that for CsFe$_2$Se$_3$ at $T$=150 K (and $T$=200 K) is 
$\sim$ 30$\%$ (and $\sim$ 0$\%$)  larger than that at  $T$=300 K.
Therefore, the present temperature dependence of XLD strongly supports the scenario of the orbital order or fluctuation in these systems even above $T_{\rm N}$.

\begin{figure*}[t!]
	\includegraphics[width=1\linewidth]{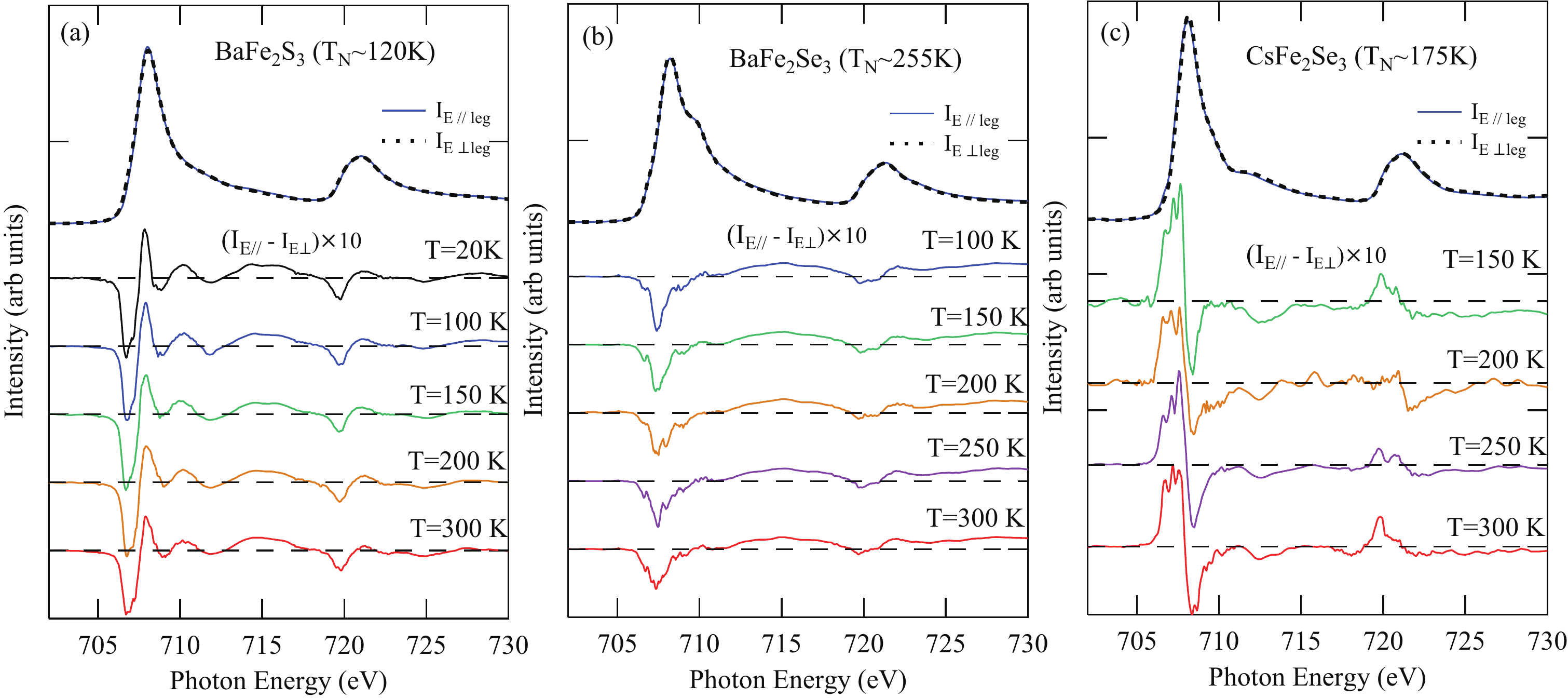}%
	\caption{(Color online)
		Temperature dependence of XLD at the Fe $L_{2,3}$ edges in the TEY modes for the different badges of crystals, (a) BaFe$_2$S$_3$, (b) BaFe$_2$Se$_3$, and (c) CsFe$_2$Se$_3$.
	}
\end{figure*}

In order to examine the energy levels in the Fe sites with unique anisotropies discussed above, polarization dependent RIXS spectra have been acquired for incident energies $h\nu_i$ across the Fe $L_3$-edge XAS spectra as indicated by the arrows in Fig. 2(a).
The data are shown in Fig. 5 for (a) BaFe$_2$S$_3$, (b) BaFe$_2$Se$_3$, and (c) CsFe$_2$Se$_3$, respectively, on an energy loss scale $\hslash\Omega　=h\nu_f-h\nu_i$, where $h\nu_f$ is the energy of outgoing photons.
Although some elastic lines are observed at zero energy loss taken with the vertical polarization, these are barely observed on the spectra with horizontal polarization owing to the experimental geometry given in the inset of Fig. 5(d).
The strong fluorescence indicated by the blue arrows are observed on the spectra with $h\nu_i>$706.8 eV and disperses from 1 eV energy loss to higher energy losses, which was ascribed by the hybridization effects between Fe 3$d$ states and Se 4$p$ states in the high-resolutional study for BaFe$_2$Se$_3$ by Monney \textit{et al.}\cite{Monney}
The fluorescence contributions shift to higher energy losses for increasing incident energies, as fluorescence in RIXS typically occurs at fixed x-ray emission energy.
Monney \textit{et al.} also suggested two Raman-like peaks labeled as A and B, not moving in energy position with variation of incident photon energy, superimposed on top of the fluorescence,
which are clearly seen in the spectra with $h\nu_i$ = 705.5 eV zoomed in Fig. 5 (d).
These two Raman-like peaks A and B are corresponding to the energy of $d$-$d$ excitations in the Fe sites [see Fig. 1(c)].
The peak energies of A roughly correspond to the magnitude of band gap and increase in going from 0.4 eV for BaFe$_2$S$_3$, 0.8 eV for BaFe$_2$Se$_3$, to 1.3 eV for CsFe$_2$Se$_3$,
which are more or less consistent with the order for the activation energies\cite{Lei,Hirata} and threshold energies of the photoemission spectroscopy.\cite{Ootsuki}

\begin{figure}[!]
	\includegraphics[width=1\linewidth]{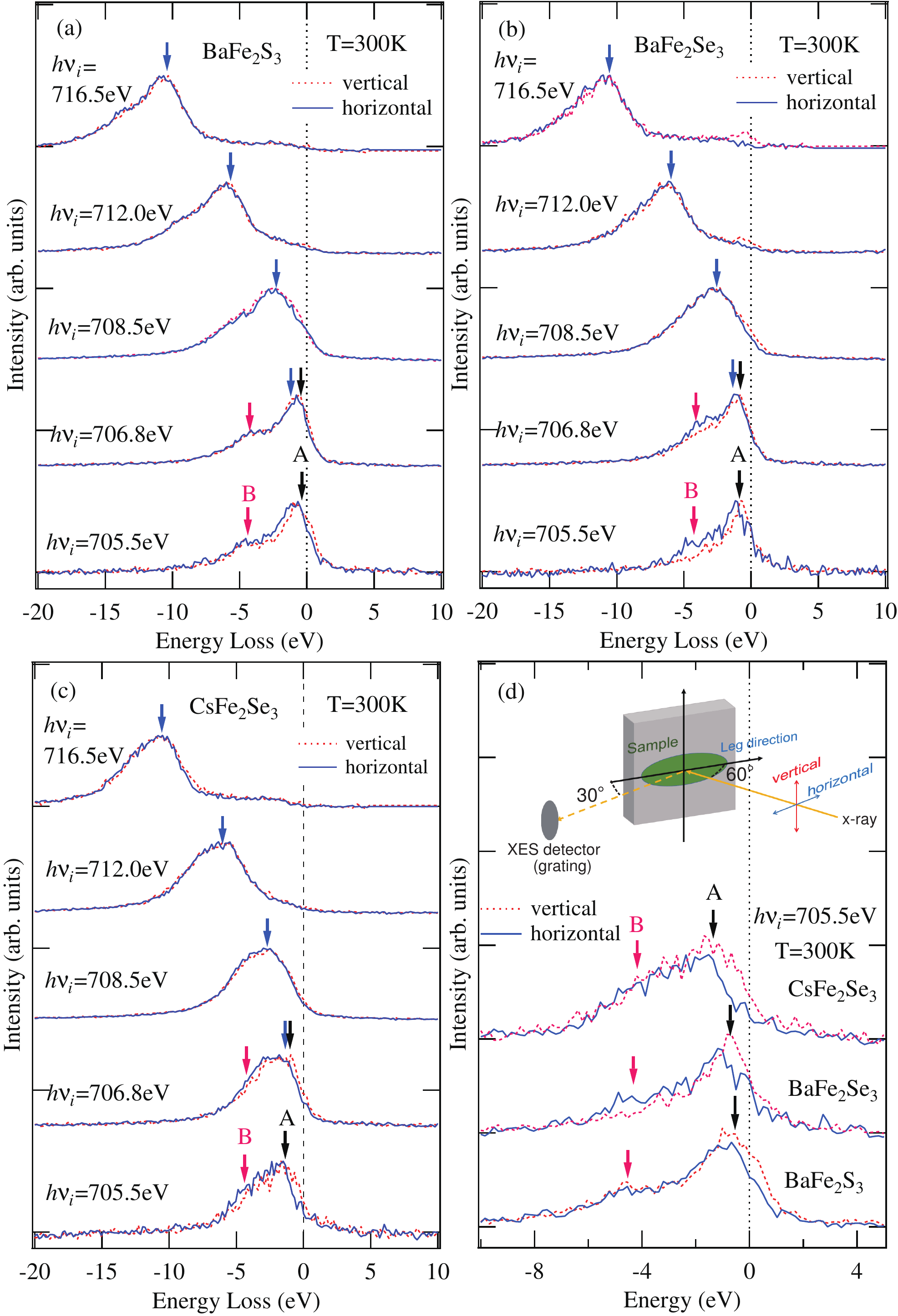}%
	\caption{(Color online)
		RIXS spectra measured with selected incident energies at the Fe $L_3$ edge for (a) BaFe$_2$S$_3$, (b) BaFe$_2$Se$_3$, and (c) CsFe$_2$Se$_3$ at $T$ = 300 K.
		The blue arrows indicate the contribution from the fluorescence.
		The black and red arrows indicate Raman-like peaks labeled as A and B. 
		(d) RIXS spectra with $h\nu$ = 705.5eV, zoomed into the low-energy-loss region.
		The inset shows the experimental geometry of RIXS.
	}
\end{figure}

The relatively large band gap observed in RIXS and orbital order along the rung direction clarified by XLD for CsFe$_2$Se$_3$ can be
explained by the molecular orbital formation between the two Fe sites of  the rung, namely ferromagnetic dimer formation.\cite{Pardo08,Mazin12}
In this scenario, the bonding orbital accommodates two electrons of the $e_g$ states in the two Fe sites across the rung in CsFe$_2$Se$_3$ and the gap is opened between the bonding and antibonding states.
Since the $d_{x^2-y^2}$ orbitals along the rung become a bonding localized state, the $d$-$d$ excitation of A on RIXS with the vertical polarization and $I_{E//{\rm leg}}$ of the pre-edges on XLD are enhanced.
On the other hand, the ferromagnetic dimers are destabilized partially in BaFe$_2$Se$_3$ and completely in BaFe$_2$S$_3$ and then $I_{E\perp{\rm leg}}$ of XLD are enhanced.
The gap sizes depend on the transfer between the dimers and therefore become very small in BaFe$_2X_3$.
This scenario seems to be consistent with a recent inelastic neutron scattering study for BaFe$_2$S$_3$.\cite{Wang}
It indicates a strong intraladder ferromagnetic exchange interaction along the rung direction, although BaFe$_2$S$_3$ still exhibits the commonly striped AF spin excitations.
In addition, the importance of Hund's rule coupling has generally been suggested in iron-based superconductors, which leads to ferromagnetic interaction between the itinerant electrons and local moments.\cite{Haule}

\section{Summary}
We have studied the electronic structures of BaFe$_2X_3$ ($X$ = S and Se) and CsFe$_2$Se$_3$ using x-ray absorption and resonant inelastic x-ray scattering spectroscopy.
XAS peak structure at the Fe $L$ edges consists of the two components in BaFe$_2X_3$, indicating that the itinerant and localized Fe 3$d$ electrons coexist.
On the other hand, the sharp peak at the Fe $L$ edges for CsFe$_2$Se$_3$ exhibit the single component accompanied with the well separated charge-transfer like satellite.
The distinct electronic anisotropies in the Fe 3$d$ states are inferred from the XLD spectra.
Different types of the orbital order or fluctuation exist in BaFe$_2X_3$ and CsFe$_2$Se$_3$ even at room temperature far above $T_{\rm N}$, which originate from the direct exchange between the $d_{3z^2-r^2}$ orbitals and molecular orbital formation bridging the rungs, respectively.
The similarity between these findings and the electronic nematic order observed in other families of the iron-based superconductors having square lattices suggests that the similar exotic phases can be realized in the quasi-one-dimensional structure.

\section*{Acknowledgements}
The authors thank Dr. D. Ootsuki and Dr. Y. Hirata for valuable discussions.
Research described in this paper was performed at the Canadian Light Source, which is supported by the Canada Foundation for Innovation, Natural Sciences and Engineering Research Council of Canada, the University of Saskatchewan, the Government of Saskatchewan, Western Economic Diversification Canada, the National Research Council Canada, and the Canadian Institutes of Health Research.
This works was supported by the Japan Society for the Promotion of Science (JSPS) of Grant-in-Aid for Young Scientists (B) (No. 16K20997) and for Scientific Research (B) (No. 16H04019).
This work was also partially supported by Ministry of Education, Culture, Sports, Science, and Technology of Japan (X-ray Free Electron Laser Priority Strategy Program) and Mitsubishi Foundation.


\end{document}